# Literature Mining System for Nutraceutical Biosynthesis: From AI Framework to Biological Insight

Xinyang Sun[1], Nipon Sarmah[1], Miao Guo[1,*]

[1]Department of Engineering, Faculty of Natural, Mathematical & Engineering Sciences, King's College London, Strand Campus, London, WC2R 2LS, UK

***Corresponding Author**
Miao Guo
E-mail address: miao.guo@kcl.ac.uk

## Abstract

The extraction of structured knowledge from scientific literature remains a major bottleneck in nutraceutical research, particularly when identifying microbial strains involved in compound biosynthesis. This study presents a domain-adapted system powered by large language models (LLMs) and guided by advanced prompt engineering techniques to automate the identification of nutraceutical-producing microbes from unstructured scientific text. By leveraging few-shot prompting and tailored query designs, the system demonstrates robust performance across multiple configurations, with DeepSeek-V3 outperforming LLaMA-2 in accuracy, especially when domain-specific strain information is included. A structured and validated dataset comprising 35 nutraceutical–strain associations was generated, spanning amino acids, fibers, phytochemicals, and vitamins. The results reveal significant microbial diversity across monoculture and co-culture systems, with dominant contributions from *Corynebacterium glutamicum*, *Escherichia coli*, and *Bacillus subtilis*, alongside emerging synthetic consortia. This AI-driven framework not only enhances the scalability and interpretability of literature mining but also provides actionable insights for microbial strain selection, synthetic biology design, and precision fermentation strategies in the production of high-value nutraceuticals.

# Introduction

Nutraceuticals are bioactive compounds derived from food sources that provide medical or health benefits beyond basic nutrition. It includes functional foods, dietary supplements, and fortified products aimed at disease prevention and health promotion[1].

The global nutraceuticals market was valued at approximately USD 712.97 billion in 2023 and is projected to grow at a CAGR of 8.4% through 2030, driven by increasing consumer interest in preventive health, clean-label ingredients, and personalised nutrition[2]. The United Kingdom is witnessing rapid expansion in this sector. According to GlobeNewswire, the UK nutraceutical market is expected to reach £8.36 billion by 2029, growing at a CAGR of 6.9% between 2024 and 2029. This growth is driven by strong consumer demand for functional foods and beverages that support immune health, gut health, and mental well-being. Popular trends include plant-based formulations, fermented products, bioactives such as probiotics and omega-3s, as well as innovations in precision fermentation for producing high-value compounds, including amino acids and peptides.

However, challenges persist in terms of regulatory alignment, scientific substantiation of health claims, and consumer awareness. Strict frameworks require rigorous evidence for nutrition and health claims, often limiting product approvals despite growing market demand[2,3]

The integration of Artificial Intelligence (AI), particularly Large Language Models (LLMs), has transformed scientific literature mining by enabling scalable and context-aware information extraction. Traditional tools such as PubTator[4], SciSpacy[5], and MetaMap[6] offered basic capabilities to perform such tasks. On the other hand, transformer-based LLMs like BioBERT[7], SciBERT[8], and PubMedBERT[9] have shown improved semantic understanding and domain adaptation by being pretrained or fine-tuned on biomedical corpora.

Prompt engineering has become an important technique in maximising the utility of LLMs[10,11]. State-of-the-art prompt engineering involves crafting input prompts that guide model behaviour to perform specific tasks, ranging from question answering and summarisation to domain-specific information extraction. Techniques have evolved beyond simple instruction-based prompts to include advanced formulations such as chain-of-thought prompting[12], which encourages the model to reason step-by-step, and few-shot prompting[11], where examples are embedded within the prompt to establish task patterns. As LLMs are deployed in increasingly complex settings, prompt engineering plays a critical role in aligning model capabilities with user intent and domain-specific requirements.

In summary, the use of LLMs provides a powerful foundation for extracting scientific knowledge in complex domains, such as the production of nutraceuticals. By leveraging prompt engineering, such systems can fill gaps in structured scientific understanding, enabling automated extraction.

A central challenge in extracting insights from scientific literature lies in the unstructured nature of the information. In domains such as nutraceutical production, important data, including microbial species, are often dispersed across text, tables, figures, and supplementary materials. The multi-modal nature of scientific outputs further compounds this fragmentation: essential experimental details may appear only in figure captions or complex tables, making them difficult to detect and interpret using traditional text-focused tools.

Beyond literature, static data sources, including both open-access repositories and curated databases, form the foundation for AI-driven discovery in the nutraceutical domain. Platforms such as PubMed[13], Europe PMC[14], and arXiv host vast corpora of scientific publications, while structured databases like FooDB[15], DSLD[16], and NCBI Taxonomy[17] offer entries for known compounds and strains.

It is therefore beneficial to have a systematic pipeline that extracts and harmonises multi-modal content, such as text parsing, before feeding it into LLMs guided by domain-specific prompt engineering. Such an approach is crucial for developing integrated, high-quality knowledge papers and datasets that can inform robust models for nutraceutical research.

Current approaches to extracting structured knowledge from scientific literature in the nutraceutical domains face several key limitations. One major gap is the lack of domain-specific tools that can accommodate the complex linguistic structures and scientific context unique to microbial fermentation and nutraceutical research. Although LLMs offer strong capabilities, they often fall short when applied without domain-adapted prompt engineering. Furthermore, the growing prevalence of multi-modal scientific data, such as textual descriptions and tables, adds another layer of complexity, limiting LLMs' effectiveness in generating comprehensive, structured knowledge from diverse data sources.

This paper addresses these challenges by introducing an LLM-based system that identifies nutraceutical compounds and detects microbial strains. A key innovation lies in the use of LLMs that are carefully guided to perform targeted extraction tasks, leveraging prompt engineering and task-specific logic to enhance reliability and domain alignment. Through these contributions, the proposed framework offers a scalable, interpretable, and domain-aware solution to accelerate literature-based discovery and knowledge curation in nutraceutical production research.

# Methodology

The study centers on the use of LLMs to extract microbial strains reported to produce nutraceutical compounds from scientific papers published on arXiv. Relevant articles were programmatically collected using the Python library *feedparser*[18] to access and parse the arXiv API. A defined set of domain-specific keyword queries related to nutraceuticals and microbial fermentation were used to retrieve candidate papers.

Given the text of an arXiv paper, the LLM was prompted to identify microbial strains that are explicitly described as producing nutraceutical compounds. The model output was constrained to strain names supported by explicit statements in the text, avoiding speculative or inferred associations.

Two LLMs with different model capacities were evaluated. LLaMA-2[19] was used in its 70-billion-parameter configuration. DeepSeek-V3[20] was employed with 671 billion total parameters. This comparison enables an analysis of how model scale influences performance on domain-specific extraction tasks. Few-shot prompting was adopted to guide the models' behaviour. Each prompt included two manually curated examples demonstrating the desired input–output format, where a short excerpt from a paper was paired with the correct strain extraction. These examples were designed to illustrate how nutraceutical production is typically described in the literature and how relevant strain names should be identified and reported. This few-shot setting was selected based on preliminary experiments which confirmed that the few-shot configuration outperformed one-shot and zero-shot variants..

A two-stage search strategy was implemented to assess the impact of query specificity on extraction accuracy. In the first stage, the model was prompted using only the nutraceutical name as the search focus, requiring the LLM to identify producing strains based solely on compound-related context. In the second stage, both the nutraceutical name and candidate strain names were included in the query, providing more explicit biological context.

Model performance was assessed using accuracy, defined as the proportion of correctly identified microbial strains relative to the total number of predicted entries. All extracted strain–compound associations were manually validated by domain experts against the original paper content, with ambiguous cases resolved through consensus review. Comparative analyses were conducted across LLM architectures, prompting strategies, and query configurations. This assessment also facilitated a qualitative characterisation of recurrent error types, including false-positive strain identifications and misinterpretation of organisms mentioned in the text but not explicitly described as nutraceutical producers.

# Results

## Evaluation of LLM and Prompt Engineering Techniques

Table 1 shows the performance of DeepSeek-V3 and LLaMA-2, evaluated across three prompt engineering strategies: few-shot, one-shot, and zero-shot. DeepSeek-V3 extracted a total of 101, 109, and 109 items in the few-shot, one-shot, and zero-shot settings, with 72, 71, and 68 correct extractions, respectively. LLaMA-2 extracted 109, 99, and 97 items across the same prompt types, with 71, 63, and 60 correct extractions. These results indicate that both models produced more correct information under few-shot prompting, suggesting that providing additional examples improves extraction accuracy. Overall, DeepSeek-V3 slightly outperformed LLaMA-2 in terms of the number of correct extractions across all prompt types.

Table 1: Model and Prompt-Engineering Result Comparison

| Model | | Few Shot | One Shot | Zero Shot |
|---|---|---|---|---|
| DeepSeek-V3 | Total | 101 | 109 | 109 |
| | Correct | 72 | 71 | 68 |
| LLaMA-2 | Total | 109 | 99 | 97 |
| | Correct | 71 | 63 | 60 |

Figure 1 compares the performance of LLaMA-2 and DeepSeek-V3 across different prompt engineering techniques, few-shot, one-shot, and zero-shot, using the percentage of correct responses as the evaluation metric. Overall, DeepSeek-V3 consistently outperforms LLaMA-2 across all prompting settings. The largest performance gap is observed in the few-shot setting, where DeepSeek-V3 achieves 71.29% accuracy compared to 65.14% for LLaMA-2, indicating a strong ability to leverage multiple in-context examples. In the one-shot scenario, both models experience a slight reduction in accuracy, though DeepSeek-V3 (65.14%) maintains a modest advantage over LLaMA-2 (63.64%). Performance further declines under the zero-shot setting, with DeepSeek-V3 attaining 62.39% accuracy and LLaMA-2 achieving 61.86%, suggesting that the absence of examples negatively impacts both models. These results highlight the importance of prompt engineering, particularly few-shot prompting, and demonstrate the superior robustness of DeepSeek-V3 across varying levels of contextual guidance.

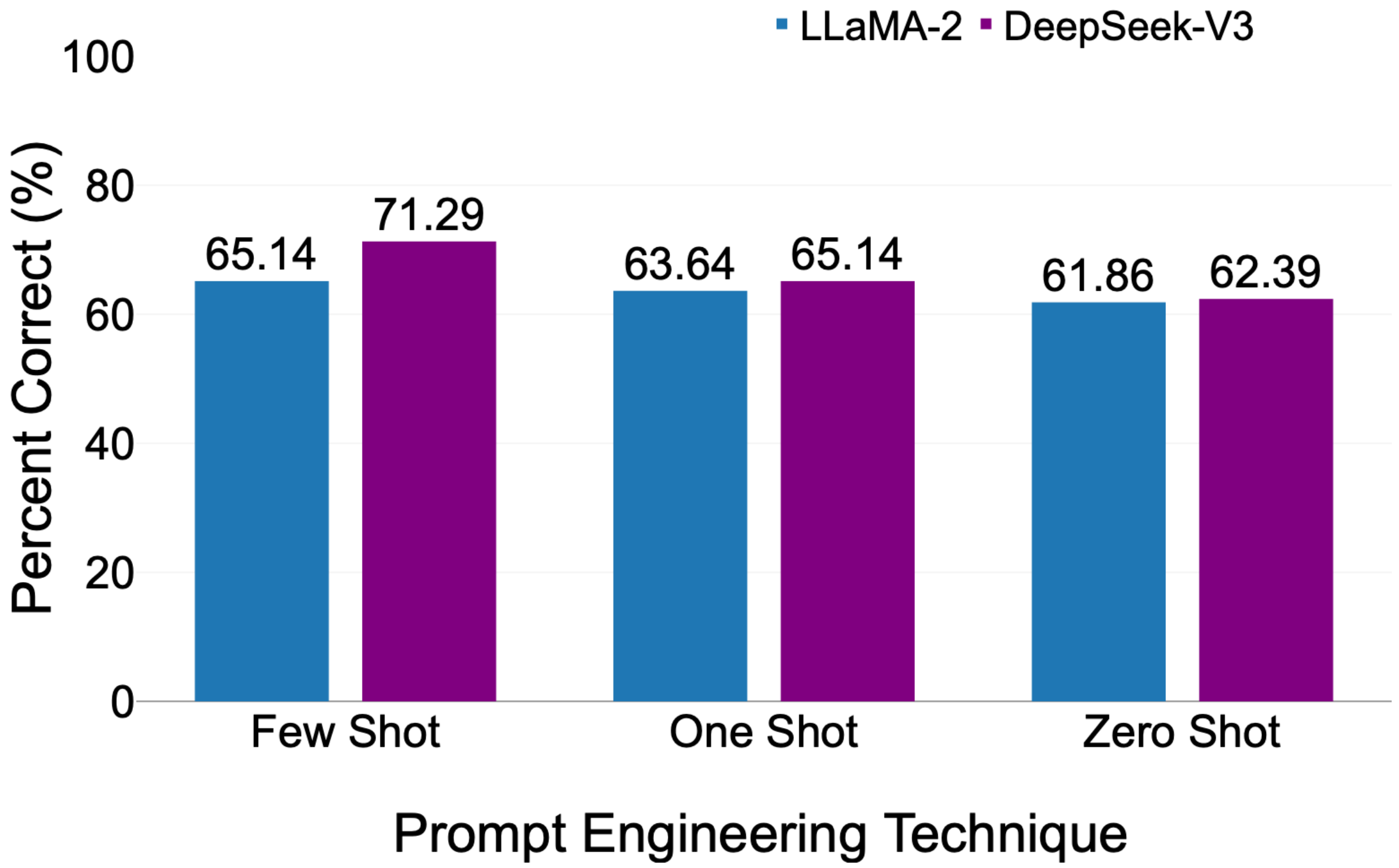

Figure 1: Model and Prompt-Engineering Result Comparison

Figure 2 illustrates the impact of incorporating strain name information into the search strategy when using the best-performing configuration, namely DeepSeek-V3 with few-shot prompting. The results demonstrate a substantial improvement in performance when strain names are explicitly included. Specifically, accuracy increases from 71.29% without strain name information to 82.76% with strain name information, representing an absolute gain of 11.47 percentage points. This marked improvement indicates that providing domain-specific contextual cues significantly enhances the model's ability to retrieve and reason over relevant information. The findings suggest that, even under an optimised prompting strategy, the inclusion of precise biological identifiers plays a critical role in improving model accuracy, highlighting the importance of structured and informative query formulation in domain-specific search and reasoning tasks.

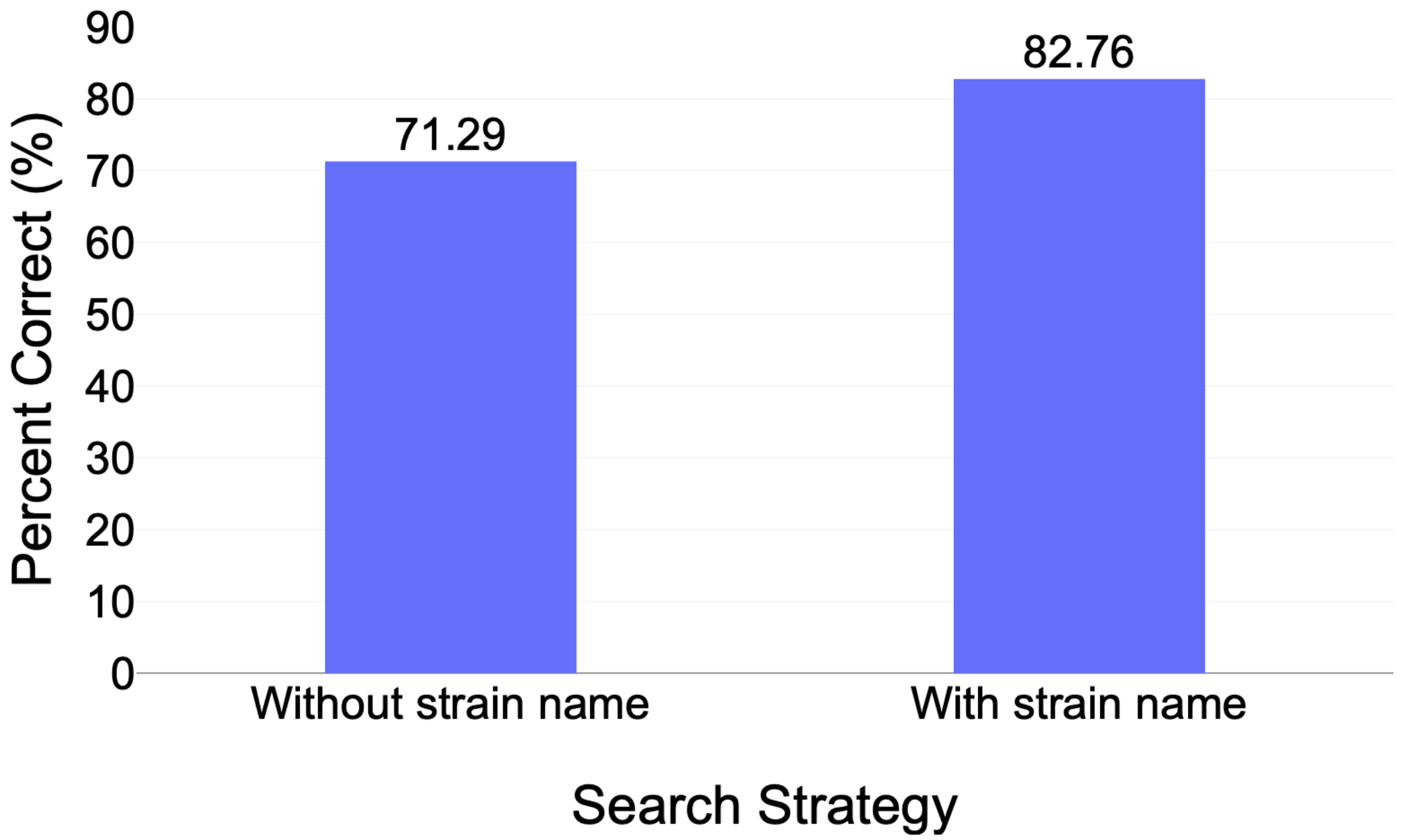

Figure 2: Search Strategy Comparison

These findings not only validate the effectiveness of few-shot prompting in domain-specific information retrieval but also demonstrate that model performance can be significantly boosted by incorporating structured biological context, a crucial insight for deploying LLMs in scientific discovery workflows.

## Domain-Level Interpretation of Extracted Results

The integration of AI-assisted literature mining with expert biological validation yielded a structured, high-confidence dataset of microbial producers for nutraceutical biosynthesis. This curated dataset captures the functional diversity of microbial fermentation systems, reflecting both classical and emerging biosynthetic routes. A total of 35 validated nutraceutical-strain associations were identified, covering 23 unique nutraceuticals and involving 31 distinct microbial strains, including both wild-type and genetically engineered organisms. Each strain-compound pair was categorised based on monoculture (single-strain systems) or co-culture (multi-strain consortia) based on the presence of single or multiple microbial producers in the validated dataset (Table 2).

Table 2: Validated nutraceutical-microbe associations derived from AI-assisted literature mining

| Category | Nutraceutical | Strain (Monoculture, Coculture) | References |
|---|---|---|---|
| Amino Acids | Alanine | *Bacillus subtilis* | [21] |
| | | *Enterococcus faecalis* | [22] |
| | | *Staphylococcus staphylolyticus* | [23] |
| | | *Staphylococcus aureus* | [24] |
| | Arginine | *Escherichia coli** | [25] |
| | Cysteine | *Salmonella typhimurium* | [26] |
| | Gamma-aminobutyric acid | *Pediococcus pentosaceus, Lacticaseibacillus rhamnosus* | [27] |
| | Glutamic acid | *Corynebacterium glutamicum* | [28] |
| | Glutamine | *Escherichia coli* | [29] |
| | Glycine | *Lactobacillus bulgaricus, Streptococcus thermophilus* | [30] |
| | Histidine | *Bacillus subtilis* | [31] |
| | | *Escherichia coli* | [32] |
| | Isoleucine | *Escherichia coli** | [33] |
| | Leucine | *Corynebacterium glutamicum* | [34] |
| | | *Sulfolobus solfataricus* | [35] |
| | Lysine | *Corynebacterium glutamicum* | [36] |
| | | *Escherichia coli** | [37] |
| | | *(Corynebacterium glutamicum, Escherichia coli)** | [38] |
| | Methionine | *Escherichia coli** | 39 |
| | Phenylalanine | *Pseudomonas putida** | 40 |
| | Serine | *Bacillus subtilis* | 41 |
| | | *Saccharomyces cerevisiae** | 42 |
| | | *Escherichia coli** | 43 |
| | Tryptophan | *Escherichia coli** | 44 |
| | Tyrosine | *Bacillus subtilis* | 45 |
| Fibers & Specialty Carbohydrates | Alginate | *Pseudomonas aeruginosa* | 46 |
| | | *Pseudomonas putida* | 47 |
| | Butyrate | *Bacteroides thetaiotaomicron, Methanobrevibacter smithii, Eubacterium rectale* | 48 |
| Phytochemicals & Plant Extracts | Cellulose | *Komagataeibacter sucrofermentans* | 49 |
| | | *Escherichia coli** | 50 |
| | Chlorophyll a | *Chlorella sorokiniana* | 51 |
| | | *Synechocystis* | 52 |
| | Gluconic acid | *Gluconobacter oxydans* | 53 |
| Vitamins | Tocopherol | *Synechocystis* | 54 |
| | Riboflavin | *Escherichia coli** | 55 |

*Genetically engineered

This domain-level interpretation provides a foundational map of current microbial biosynthetic capacities with direct relevance to industrial fermentation. It highlights not only strain diversity but also the strategic use of engineered microbes and consortia for enhanced production efficiency. These insights provide tangible value for applications in metabolic engineering, synthetic biology, and precision fermentation, particularly in the context of nutraceutical manufacturing.

## Nutraceutical Classes & Biosynthetic Value

We first examine the diversity of nutraceutical classes identified through literature mining. Among all identified categories, Amino Acids emerged as the most prominent class, representing over 60% of the validated nutraceutical-strain associations. These include essential amino acids such as lysine, leucine, methionine, and isoleucine, along with functional compounds like GABA (gamma-aminobutyric acid), which has gained popularity in functional foods due to its neuroactive and anti-stress properties[56]. This dominance reflects both the industrial maturity of amino acid fermentation and the deep metabolic engineering knowledge base associated with microbial hosts such as *Corynebacterium glutamicum*[28,34,36], *Escherichia coli* [25,32,33,37,39,43,44], and *Bacillus subtilis*[21,31,41,45].

Beyond amino acids, the dataset includes compounds from Fibers & Speciality Carbohydrates, notably butyrate and alginate. These compounds play critical roles in gut health modulation[48] and textural or stabilising functions in food systems[46,47]. Notably, butyrate was validated as a product of a complex co-culture involving *Bacteroides thetaiotaomicron*, *Methanobrevibacter smithii*, and *Eubacterium rectale*, mimicking gut microbial metabolism[48].

The category of Phytochemicals & Plant Extracts includes compounds such as chlorophyll a, cellulose, and gluconic acid, typically associated with plant or algal systems but here produced by microbial hosts such as *Synechocystis*[52], *Komagataeibacter sucrofermentans*[49], and *Gluconobacter oxydans*[53]. Their presence in the dataset underscores the increasing importance of microbial platforms in replacing plant extraction with sustainable, fermentation-based alternatives.

Lastly, Vitamins, though fewer in number, demonstrate the high-value potential of microbial biosynthesis. Tocopherol (Vitamin E) and riboflavin (Vitamin B2) were linked to *Synechocystis*[54] and engineered *Escherichia coli*[55], respectively, reflecting advances in vitamin production using recombinant systems.

This functional and taxonomic diversity of compounds highlights the biosynthetic flexibility of microbial hosts. It validates the literature-mining system's ability to identify both industrially relevant and emerging nutraceutical targets across chemical classes.

## Microbial Strains

The validated dataset comprises a diverse array of microbial producers, including Gram-positive and Gram-negative bacteria, actinobacteria, archaea, cyanobacteria, and yeast, reflecting both classical fermentation organisms and emerging biotechnological chassis. These microbes are well-documented for their fermentative robustness, genetic tractability, and historical use in nutraceutical bioproduction. The microbial strains identified can be effectively categorised by their application in monoculture or co-culture systems, revealing distinct biological strategies for nutraceutical production. The distribution across these two formats provides critical insight into current microbial fermentation strategies for nutraceutical production and informs both strain engineering and bioprocess design decisions for future development. A circular taxonomic representation of all validated microbial producers, categorised by nutraceutical class, compound type, and culture format (monoculture vs. co-culture), is presented in Figure 3.

### *Monoculture*

Next, we examine the microbial taxa involved in production, with a focus on monoculture systems. Monoculture fermentation employing a single microbial strain for biosynthesis remains the dominant strategy in the validated dataset and the broader industrial landscape. It is favoured for its high degree of process control, predictable metabolic outputs, and scalability in commercial fermentation systems[57,58]. The extracted literature and validation efforts highlight a strong reliance on well-characterised, genetically tractable microbial workhorses.

The Gram-positive bacterium *Corynebacterium glutamicum* is prominently featured as a monoculture producer of several high-value amino acids, including glutamic acid[28], leucine[34], and lysine[36], confirming its longstanding role as a premier microbial cell factory for amino acid production. Its robustness under minimal media conditions and low by-product formation makes it especially attractive for high-titer production.

*Escherichia coli*, both wild-type and genetically engineered strains, appears frequently in association with the production of arginine[25], serine[43], riboflavin[55], and tryptophan[44]. Its rapid growth kinetics, rich toolbox of synthetic biology parts, and well-mapped metabolic network enable fine-tuned pathway engineering for the biosynthesis of nutraceuticals.

*Bacillus subtilis* also plays a significant role, with confirmed monoculture production of alanine[21], serine[41], and tyrosine[45], further reinforcing its status as a GRAS (Generally Recognised as Safe) organism with broad applicability in food-grade biosynthesis.

In the eukaryotic domain, *Saccharomyces cerevisiae* was identified as a monoculture producer of serine[42], showcasing its metabolic versatility beyond its traditional use in ethanol production.

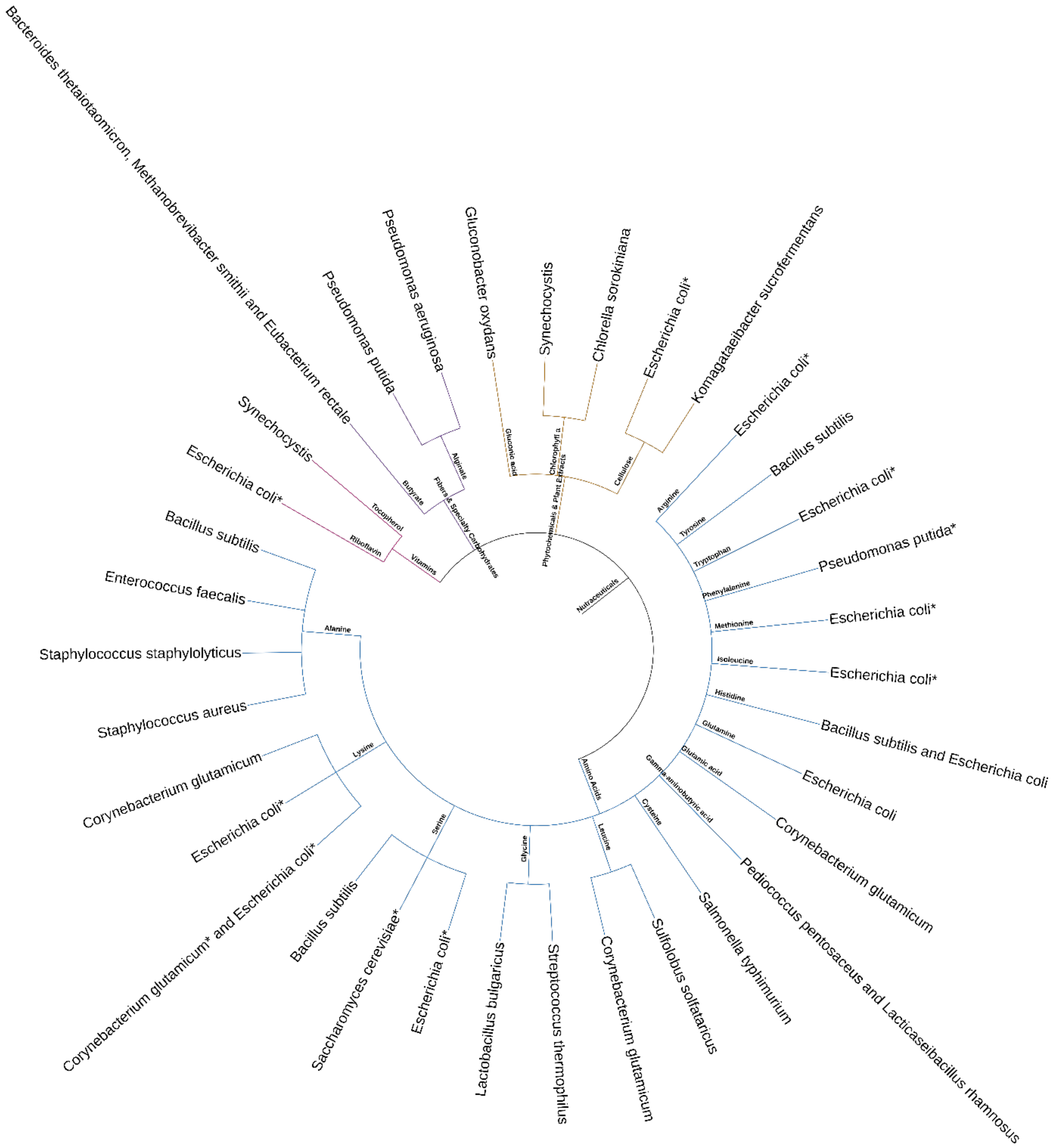

Figure 3: Circular taxonomic tree of validated microbial producers of nutraceutical compounds. Genetically engineered strains are annotated with an asterisk (*)

The recurrence of these strains underscores their metabolic robustness, engineering accessibility, and regulatory familiarity, making them the preferred hosts for controlled, single-strain fermentation systems in the production of nutraceuticals.

### *Co-culture*

Finally, we turn to co-culture strategies, which demonstrate synergistic biosynthetic pathways. Co-culture fermentation systems, though fewer in number compared to monocultures, present a distinct biosynthetic strategy that leverages metabolic interactions between multiple microbial strains. These systems introduce functional complexity and enable forms of metabolic synergy that are often unattainable in single-strain fermentations. Through division of labour, cross-feeding of intermediates, and ecological cooperation, co-cultures facilitate the synthesis of nutraceutical compounds that require multi-step biochemical conversions or tightly regulated environmental conditions[59,60].

A validated example is the production of gamma-aminobutyric acid (GABA) via a co-culture of *Pediococcus pentosaceus* and *Lacticaseibacillus rhamnosus*. Combining *Pediococcus pentosaceus* (a potent GABA producer) with *Lactobacillus namurensis* (an acid-regulating strain) significantly enhances GABA accumulation in Nham, yielding a novel fermented pork sausage with high GABA content, improved safety, and superior sensory acceptability[27]. Similarly, a complex tri-culture involving Bacteroides thetaiotaomicron, *Methanobrevibacter smithii*, and *Eubacterium rectale* was linked to butyrate production, mimicking syntrophic interactions found in the human gut microbiome. In this configuration, the strains perform complementary roles in carbohydrate degradation, hydrogen consumption, and short-chain fatty acid synthesis, collectively achieving a metabolic outcome not possible by any one strain alone[48].

Another instance of co-culture involves Lactobacillus bulgaricus and *Streptococcus thermophilus*, which have been co-fermented to produce glycine, building upon well-established interactions in dairy fermentation systems[30]. Furthermore, lysine biosynthesis has been achieved through a genetically engineered dual-strain system comprising *Corynebacterium glutamicum* and *Escherichia coli*, allowing for distributed pathway engineering and enhanced metabolic efficiency[38].

These co-culture configurations highlight the growing interest in microbiome-inspired bioproduction and the rational design of synthetic consortia. Importantly, many of these systems operate without the need for genetic engineering, relying instead on native metabolic compatibility and ecological cooperation. As such, co-culture strategies offer a promising avenue for the sustainable, modular, and adaptive biosynthesis of complex nutraceuticals, particularly in cases where single-strain systems are limited by metabolic burden or regulatory constraints.

Together, these microbial strategies reflect the evolving landscape of microbial fermentation, from robust monocultures to flexible, functionally layered co-cultures, paving the way for expanded industrial implementation. These insights underscore the broader potential of AI-guided literature mining not only as a discovery tool but also as a catalyst for innovation in synthetic biology and fermentation-based nutraceutical development.

## Use of extracted information

The extracted results offer a data-driven foundation for strain selection, bioprocess design, and ingredient sourcing in nutraceutical development. Industrial players seeking robustness and yield will find actionable leads among the monoculture entries, while synthetic ecologists and systems biologists may explore the co-culture data for insights into natural metabolic cooperation.

The structured knowledge base generated through the integration of AI-assisted mining and expert biological validation demonstrates immediate utility across multiple stages of nutraceutical research and development. Foremost, it serves as a decision-support system for strain selection and bioprocess design, enabling researchers to bypass the need for an extensive manual literature review. For instance, a metabolic engineer seeking a natural producer of alginate can directly identify *Pseudomonas aeruginosa* from the dataset, streamlining early-stage strain screening. Similarly, the recurrent association of *Corynebacterium glutamicum* with multiple amino acids provides empirical support for its continued use as a reliable chassis for amino acid biosynthesis.

Beyond individual strain identification, the dataset enables comparative analyses and trend mapping. The frequent appearance of *Escherichia coli* across multiple nutraceutical categories reinforces its role as a versatile, general-purpose microbial chassis, amenable to genetic engineering and applicable across diverse biosynthetic pathways. In contrast, more specialised associations such as *Komagataeibacter sucrofermentans* with bacterial cellulose production illustrate the dataset's capacity to capture niche-specific biosynthetic expertise within lesser-studied taxa. These associations may inform not only organism selection but also the rational design of engineered pathways and chassis strain diversification based on taxonomic or metabolic similarity.

Importantly, the curated co-culture entries provide a structured foundation for exploring synthetic microbial consortia. The tri-culture system validated for butyrate production involving *Bacteroides thetaiotaomicron*, *Methanobrevibacter smithii*, and *Eubacterium rectale* serves as a representative model for investigating syntrophic interactions, metabolite cross-feeding, and community-level metabolic dynamics. These insights are critical for translating laboratory-scale co-culture systems into stable, industrially viable fermentation platforms.

In summary, the extracted and validated dataset transforms dispersed scientific evidence into a consolidated, actionable knowledge base, accelerating the early stages of nutraceutical R&D. It supports hypothesis generation, chassis prioritisation, and model construction for both traditional monoculture-based bioproduction and more advanced, ecology-inspired co-culture systems.

# Conclusion

This study demonstrates the potential of LLMs, combined with prompt engineering, to extract structured biological knowledge from unstructured scientific literature in the nutraceutical domain. By automating the identification of microbial strains associated with compound biosynthesis, the proposed system offers a scalable and domain-aware tool for accelerating literature-based discovery.

A comparative evaluation of LLaMA-2 and DeepSeek-V3 revealed that both model architecture and prompting strategy significantly influence extraction accuracy. DeepSeek-V3 consistently outperformed LLaMA-2 across all scenarios, with the largest gains observed under few-shot prompting, underscoring the model's ability to effectively leverage contextual examples. Performance declined in one-shot and zero-shot settings, reinforcing the importance of contextual guidance.

Incorporating domain-specific information, particularly microbial strain names, into the prompt structure yielded an additional over 11% improvement in accuracy, reinforcing the value of precise query formulation and prompt design, even when deploying highly capable foundation models.

Overall, these findings demonstrate that optimal performance in LLM-based literature mining emerges from the synergy between model capacity, strategic prompt engineering, and domain-informed context. This approach not only enhances extraction accuracy but also establishes a foundation for broader applications in bioinformatics, metabolic engineering, and synthetic biology. Future work could explore the integration of multi-modal sources (such as tabular data, figures, and supplementary datasets), as well as extend the framework to additional classes of high-value bioproducts and pathway discovery tasks.